\documentclass[aps,prl,twocolumn,groupedaddress]{revtex4}

\usepackage{graphicx}

\usepackage{amssymb,amsmath}

\usepackage{pstricks}
\usepackage{pst-plot}

\newcommand{\abs}[1]{\lvert#1\rvert}

\begin{document}

\title[Necessary Conditions for Traveling Wave Solutions]
{Exact and Asymptotic
Conditions on
Traveling Wave Solutions of the Navier-Stokes Equations}

\author{Y. Charles Li}

\address{Department of Mathematics, University of Missouri, 
Columbia, MO 65211, USA}

\email{liyan@missouri.edu}

\thanks{YCL is partly supported by a DoE grant 0009527.}

\author{Divakar Viswanath}

\address{Department of Mathematics, University of Michigan, Ann Arbor,
MI 48109, USA}

\email{divakar@umich.edu}

\thanks{DV is partly supported by NSF grants
DMS-0407110 and DMS-0715510.}
\date{}



\begin{abstract}
We derive necessary conditions that traveling wave solutions of the
Navier-Stokes equations must satisfy in the pipe, Couette, and channel
flow geometries. Some conditions are exact and must hold for any
traveling wave solution irrespective of the Reynolds number
($Re$). Other conditions are asymptotic in the limit
$Re\rightarrow\infty$.  The exact conditions are likely to be useful
tools in the study of transitional structures. For the pipe flow
geometry, we give computations up to $Re=100000$ showing the
connection of our asymptotic conditions to critical layers that
accompany vortex structures at high $Re$.
\end{abstract}

\maketitle

Turbulence is sometimes said to be the last unsolved problem of classical
physics. However, in a sense it is a fully solved problem,
since we know with near certainty that the Navier-Stokes equations
(NSE), along with the no-slip boundary conditions \cite{LBS2007}, are
an excellent physical model for all the phenomena associated with
turbulence and transition.

Although the physical model is known, figuring out the solutions of
the NSE has proved to be a very difficult problem, computationally or
analytically. To illustrate this difficulty, we note that while useful
models of the structural strength of an entire aircraft can be
built using $1$ to $100$ million elements, it takes more than a
billion grid points to completely resolve the flow in a cubical volume
of extent an inch${}^3$ at the surface of a car moving at around $60$
\hbox{m.p.h.}
\cite{BK1976,HJ2006,MVNPC2006}. 

The Navier-Stokes equations, which model the evolution of the velocity
field ${\bf u}$ of an incompressible fluid, are $\partial {\bf
u}/\partial t + ({\bf u}.\nabla){\bf u} = -\nabla p + (1/Re)\triangle
{\bf u}$, where the velocity field ${\bf u}$ must satisfy the
incompressibility constraint $\nabla.{\bf u}=0$.  There is no explicit
equation for evolving the pressure $p$, and $Re$ denotes the Reynolds
number.

Traveling wave solutions of the form ${\bf u}({\bf x},t) = \tilde{\bf
u}(x-{\bf c}t)$ form our main topic. As the motion of turbulent fluids
is characterized by disordered and intermittent fluctuations about a
mean, the significance of traveling wave solutions  may
seem limited.  Indeed, one has to look at more complicated solutions
to begin to understand turbulent fluctuations
\cite{Viswanath2007}. However, there is gathering evidence that
traveling wave solutions may help understand certain coherent
structures in transitional pipe flow \cite{SEV2007,WillisK2008}.

Further, using certain
lower-branch traveling waves, we can exhibit {\it critical layers} at
high $Re$ 
\cite{Viswanath2009,WGW2007}
that are far beyond the reach of ordinary direct numerical
simulation. The Orr-Sommerfeld equation, which governs the propagation
of infinitesimal normal mode perturbations of a base flow, is singular
at $Re=\infty$, which is the origin of the theory of critical layers
\cite{Maslowe1981, Maslowe1986}. 
The ability to compute critical layers in fully resolved numerical
solutions of the NSE could be significant, as 
critical layers occur in many important situations
\cite{Maslowe1986}. The gigantic trailing
vortices that escape from the boundary layers of airplanes during
take-off may develop critical layers \cite{MN2008}. So could vortices
shed by wind turbines, with possible implications for the optimal
arrangement of turbines in a wind farm.

Many linearly unstable (and non-laminar) traveling wave solutions and 
equilibria (which are special cases with ${\bf c} = 0$)
of Couette \cite{Nagata1990,Nagata1997}, channel
\cite{Waleffe1998, Waleffe2003}, and pipe \cite{FE2003, WK2004}
geometries have been computed. A notably systematic and extensive
effort is due to Gibson and others \cite{GHC2008, GHC2009}. The
exact conditions we derive must hold for the velocity fields of
all these solutions. In addition, analogous conditions must hold
for periodic solutions and relative periodic solutions that travel
only in the streamwise direction.

The mere existence of traveling wave solutions does not imply their
relevance to phenomena as they manifest themselves in Nature and in
technology.  However, efforts to connect traveling waves computed in
short pipes to puffs have been partially successful \cite{SEV2007,
WillisK2008}. Puffs are transitional structures observed in pipes
around $Re=2000$ that are approximately $20$ pipe diameters in axial
length and which travel with a speed that is somewhat less than the
mean streamwise velocity \cite{PM2006}. Tantalizingly, there are
hints that an entire puff may correspond to a traveling wave or some
such solution of pipe flow \cite{WillisK2009}. Our exact conditions
will be helpful in the investigation of that possibility.


We now turn to the derivation of the exact conditions.  In the
velocity field ${\bf u}=(u,v,w)$ of the NSE, $u$, $v$ and $w$ are the
streamwise (coordinate axis $x$), wall-normal ($y$), and spanwise
($z$) components, respectively, for the rectangular Couette and
channel geometries. In the case of pipe flow, $u$, $v$ and $w$ are the
radial ($r$), polar ($\theta$) and streamwise (or axial) ($z$)
components, respectively.

In plane Couette flow, the walls at $y=\pm 1$ move in the $x$
direction with speeds equal to $\pm 1$. The boundary conditions in the
streamwise and spanwise directions are periodic, with the periods
taken to be $2\pi \Lambda_x$ and $2\pi \Lambda_z$, respectively.  For
pipe flow, we assume the axial or streamwise boundary condition to be
periodic with period $2\pi\Lambda$. The walls are no slip in all
cases. The derivations are given mainly for the plane Couette flow
geometry.

Let ${\bf u}({\bf x}, t) = \tilde{\bf u}({\bf x}-{\bf c}t)$ be a traveling
wave solution of plane Couette flow. We assume ${\bf c} = (c, 0, 0)$
so that the traveling wave moves in the streamwise direction only.  If
$\tilde{\bf u} = (u, v, w)$, the streamwise or $x$ component of the
NSE gives
\begin{equation}
-c {\partial_x u} + 
\bigl(u{\partial_x u} + v{\partial_y u}
+ w {\partial_z u}\bigr) = -{\partial_x p}
+ {c_p\over Re} + {1\over Re}\triangle u.
\label{eqn-1}
\end{equation}
Here $c_p/Re$ gives the pressure gradient in the streamwise directions,
with $c_p=0$ for plane Couette flow and $c_p > 0$ for channel flow.
Let $U(y,z)$ denote  $(2\pi\Lambda_x)^{-1}\int_0^{2\pi\Lambda_x}u(x,y,z)\, dx$,
which is the mean streamwise component of $u$. From \eqref{eqn-1}, we get
\begin{equation}
\overline{\bigl(u{\partial_x u} + v{\partial_y u}
+ w {\partial_z u}\bigr)} = {c_p + \triangle U\over Re},
\label{eqn-2}
\end{equation}
where the overline denotes streamwise averaging.
At the walls $y=\pm 1$, ${\partial_x u}=0$
and $v=w=0$ because of no-slip. For the same reason, 
${\partial_{zz} U} = 0$ at the walls. Therefore,
\begin{equation}
c_p + {\partial_{yy} U} = 0
\label{eqn-3} 
\end{equation}
must hold at the walls.

As the velocity field $\tilde{\bf u}$ has zero divergence, we may rewrite
\eqref{eqn-2} as 
\begin{equation}
\overline{\nabla\cdot(u^2, uv, u w)} 
= {\partial_y \overline{u v}}
+ {\partial_z \overline{u w}} 
= {c_p + \triangle U\over Re}.
\label{eqn-4}
\end{equation}
If \eqref{eqn-4} is integrated over the cross-section, Green's theorem
applies to the expression in the middle of \eqref{eqn-4}. 
The integral of the middle term must be zero
because $v = 0$ at the walls and
$u w$ is periodic in $z$. Thus we have
\begin{equation}
\int_0^{2\pi \Lambda_z} \int_{-1}^{+1} (\triangle U + c_p)\, dy\, dz = 0.
\label{eqn-5}
\end{equation}
The derivation of the necessary conditions \eqref{eqn-3} and \eqref{eqn-5}
applies to channel flow with no change. However, $c_p \neq 0$ for channel flow.

The conditions \eqref{eqn-3} and \eqref{eqn-5} must be satisfied by all
traveling wave solutions of plane Couette flow or channel flow, whose
wave speed vector ${\bf c}$ only has a streamwise component. Indeed,
those conditions must be satisfied by all periodic solutions ${\bf u}({\bf x},
t) = {\bf u}({\bf x}, t+T)$, $T$ being the period, or relative periodic
solutions ${\bf u}({\bf x},
t) = {\bf u}({\bf x+s}, t+T)$ if the shift ${\bf s}$ only has a streamwise
component. To form $U$ in those instances, one must average both over 
a single period and in the streamwise direction as a simple modification
of our derivation will show.

For the case of pipe flow, let ${\bf c} = (0,0,c)$ so that the
traveling wave travels in the streamwise direction only.  Let $W(r,\theta) = (2\pi\Lambda)^{-1} \int_0^{2\pi\Lambda}
w(r,\theta, z)\, dz$ be the mean streamwise velocity. The analogue of
\eqref{eqn-3} requires
\begin{equation}
c_p + \partial_{r}^2 W + {\partial_r W \over r} = 0
\label{eqn-6}
\end{equation}
at all points on the circumference.
If we assume the pipe radius to be $1$, the analogue of \eqref{eqn-5}
is
\begin{equation}
\int_0^{2\pi} \int_0^1 (\triangle W + c_p) r\,dr\,d\theta = 0.
\label{eqn-7}
\end{equation}
The derivation of the necessary conditions \eqref{eqn-6} and \eqref{eqn-7}
for pipe flow is similar to that of their Couette analogues.

Traveling waves normally arise from saddle-node bifurcations with
increasing $Re$ \cite{Nagata1990, Waleffe1998, Schmiegel1999, FE2003,
WK2004}. The branch corresponding to lower energy dissipation is
called the lower branch. We will now derive certain scalings with
respect to increasing $Re$ that are characteristic of the lower branch
families.

In the case of plane Couette flow or channel flow, if a traveling wave
solution is given by $\tilde{\bf u}({\bf x}-{\bf c} t)$, the velocity
field $\tilde{\bf u}({\bf x})$ can be decomposed as 
\begin{equation}
{\bf u}_0(y,z) +
\sum_{n=1}^\infty \bigl({\bf u}_n(y,z)
\exp(i\alpha x) + \text{c.c}\bigr),
\label{eqn-8}
\end{equation}
 where $\alpha = 1/\Lambda_x$. We take
${\bf u}_0 = (U, v_0, w_0)$ and ${\bf u}_i = (u_i, v_i, w_i)$ for
$i \geq 1$. For pipe flow, the decomposition analogous to \eqref{eqn-8}
is given by  ${\bf u}_0(r,\theta) +
\sum_{n=1}^\infty \bigl({\bf u}_n(r,\theta)
\exp(i\alpha z) + \text{c.c}\bigr)$,
with $\alpha = 1/\Lambda$. We take ${\bf u}_i = (u_i, v_i, w_i)$ 
for $i\geq 1$ as
for Couette flow, but ${\bf u}_0 = (u_0, v_0, W)$ for pipe flow.

The scalings of the lower branch family that are known or that will be
derived apply to the mean streamwise velocity ($U$ or $W$), or the
rolls ($(v_0,w_0)$ or $(u_0, v_0)$), or the magnitude of modes such as
${\bf u}_1$. It is an empirical fact (but see \cite{Waleffe1997,
WK2004,LHR1994}) that the rolls and the $\bf{u}_1$ mode diminish in
magnitude approximately at the rate $Re^{-1}$.  Higher modes with
$n>1$ appear to diminish even faster. The derivations assume these
scalings.  In addition, the dissipation rate of the lower branch
families decrease with increasing $Re$, assuming that the dissipation
rate of the laminar solution is normalized to be $1$
\cite{Viswanath2009,WGW2007}.

The wall-normal or $y$ component of the $n=1$ mode of the NSE gives
\begin{equation}
i\alpha(U- c) v_1 = -\partial_y p_1 + Re^{-1}
(-\alpha^2 v_1 + \partial_y^2 v_1 + \partial_z^2 v_1) + \cdots
\label{eqn-9}
\end{equation}
for plane Couette or channel flow. The first neglected terms in
\eqref{eqn-9} are $-v_0 \partial_y v_1 - w_0 \partial_z v_1
- v_1 \partial_y v_0 - w_1 \partial_z v_0$. The analogous equation for
the radial component of pipe flow is $i\alpha(W-c) u_1 = -\partial_r
p_1 + Re^{-1} \triangle_r u_1$, where $\triangle_r$ corresponds to the
usual form of the Laplacian in the radial component of the NSE.
Terms such as $-v_1 \partial_\theta v_0/r$ are neglected.

\begin{figure*}
\begin{center}
\includegraphics[scale=0.3]{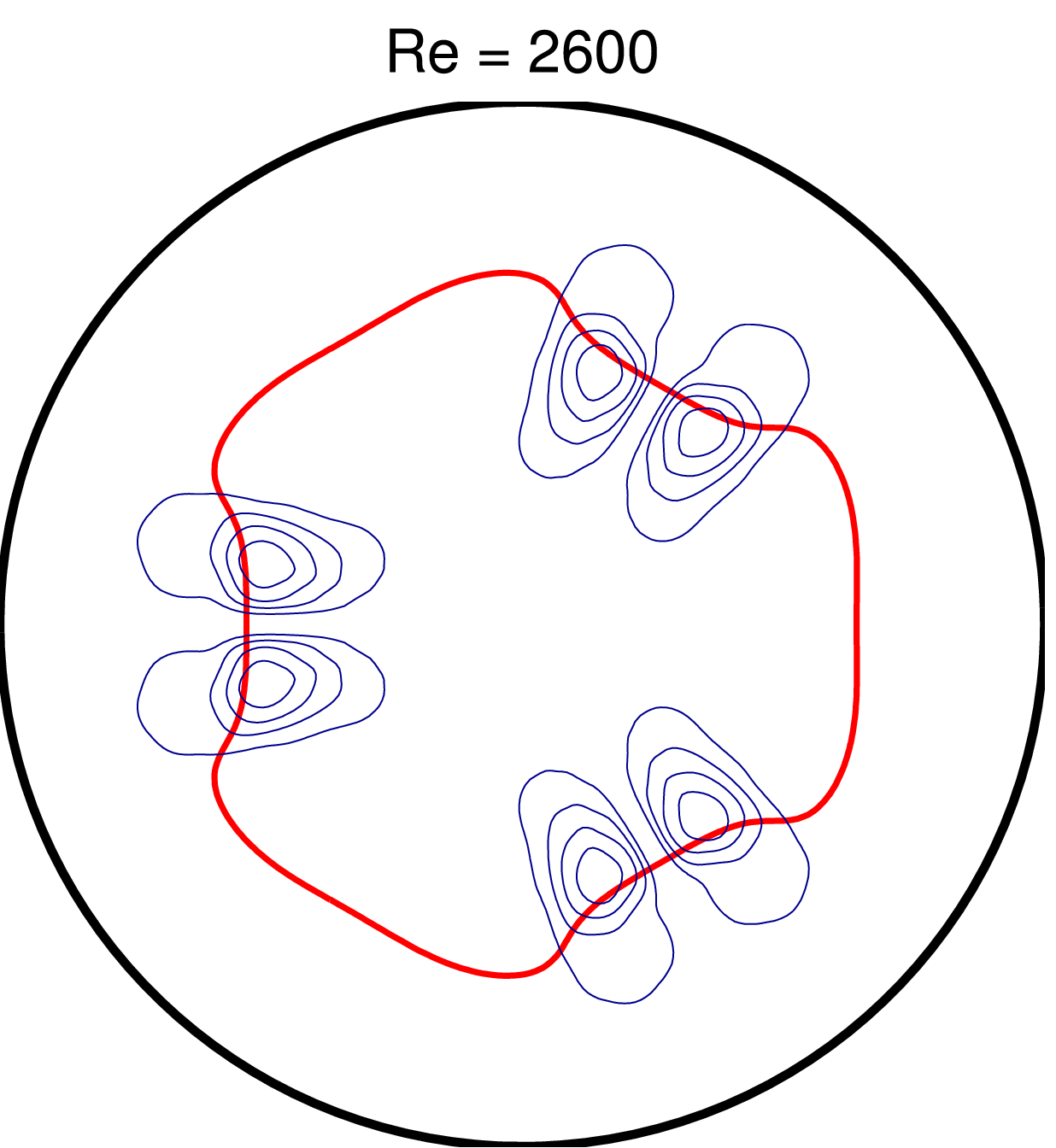}\hskip 1.5cm
\includegraphics[scale=0.3]{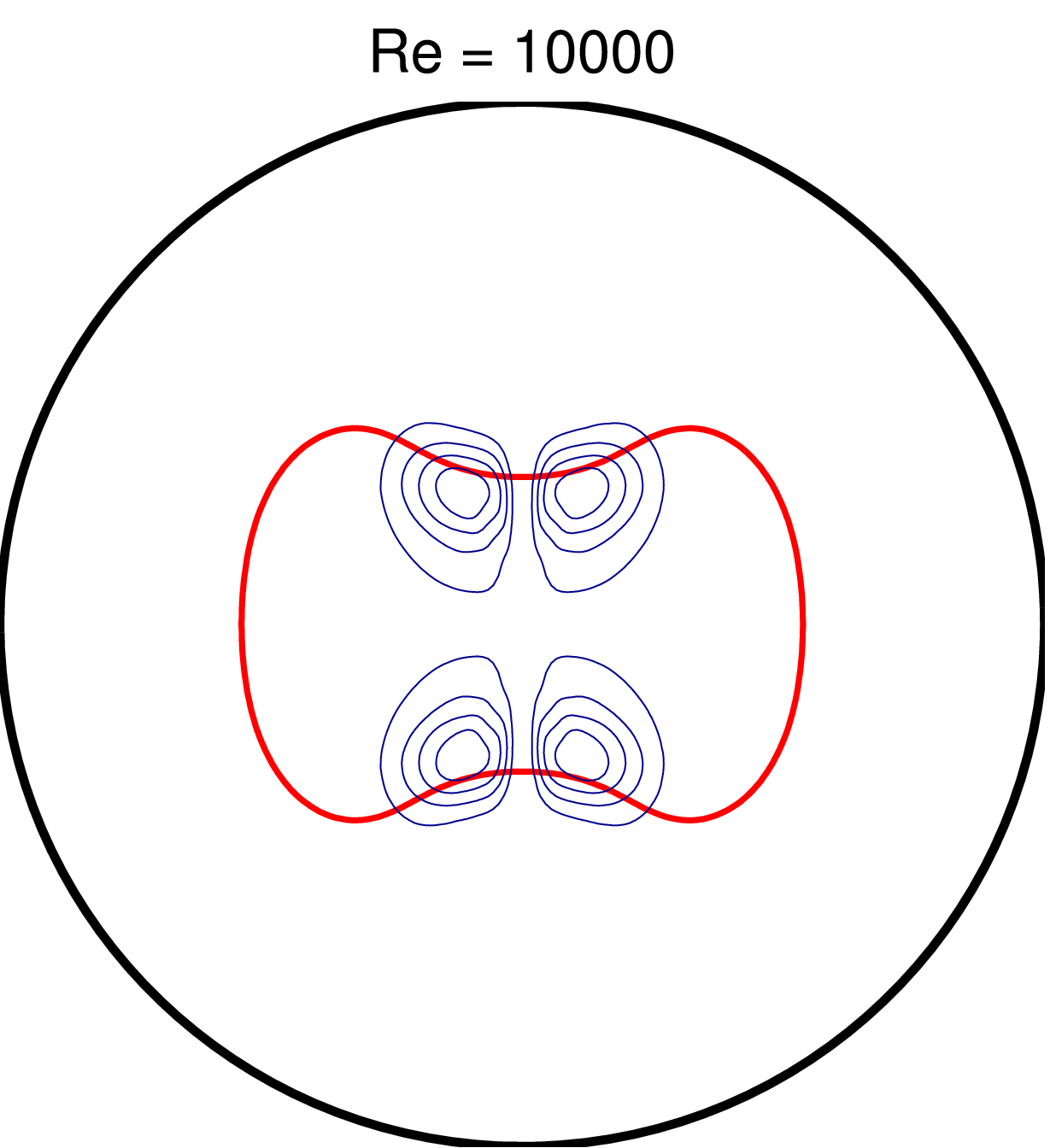}\hskip 1.5cm
\includegraphics[scale=0.3]{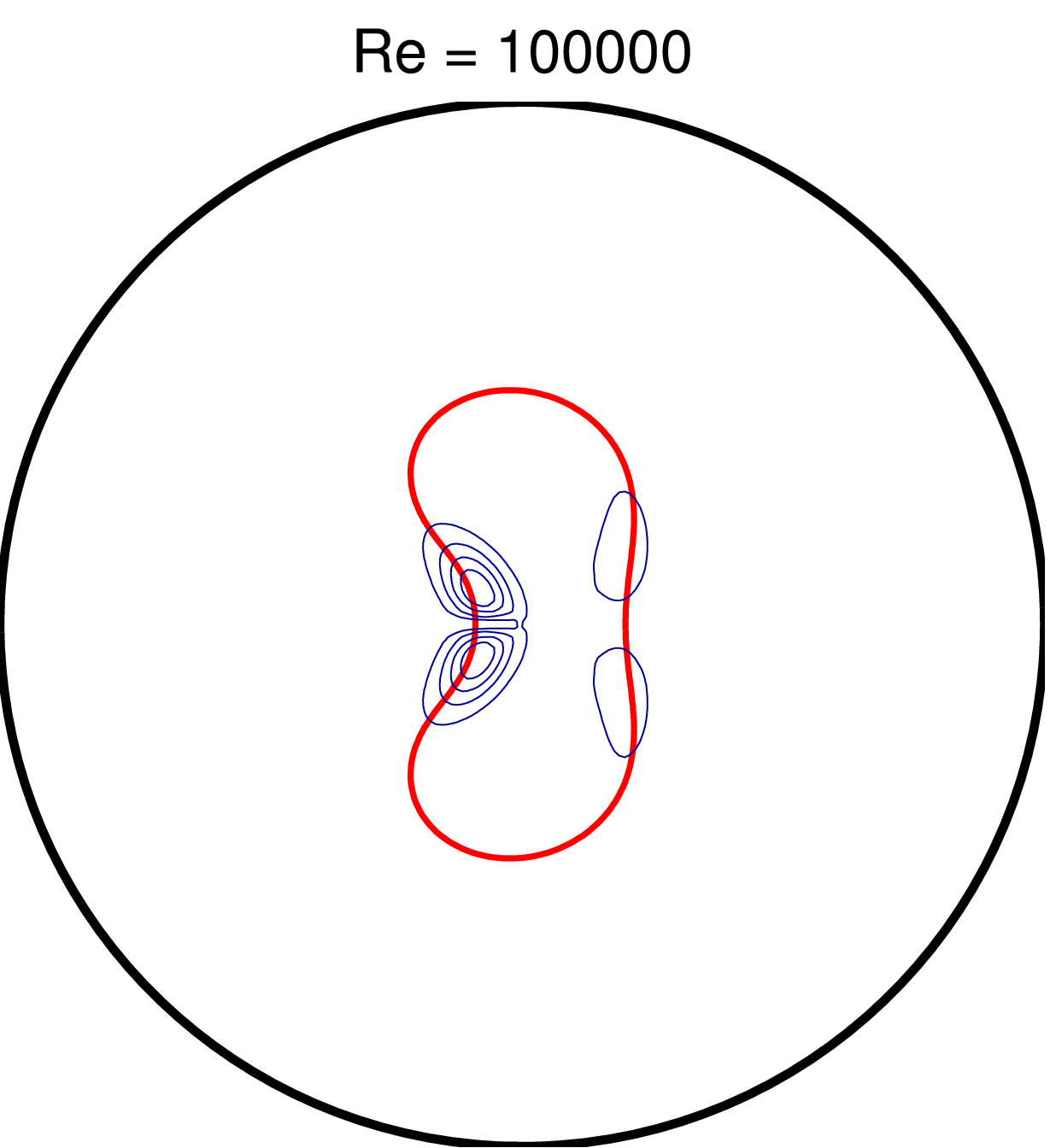}\\
\includegraphics[scale=0.3]{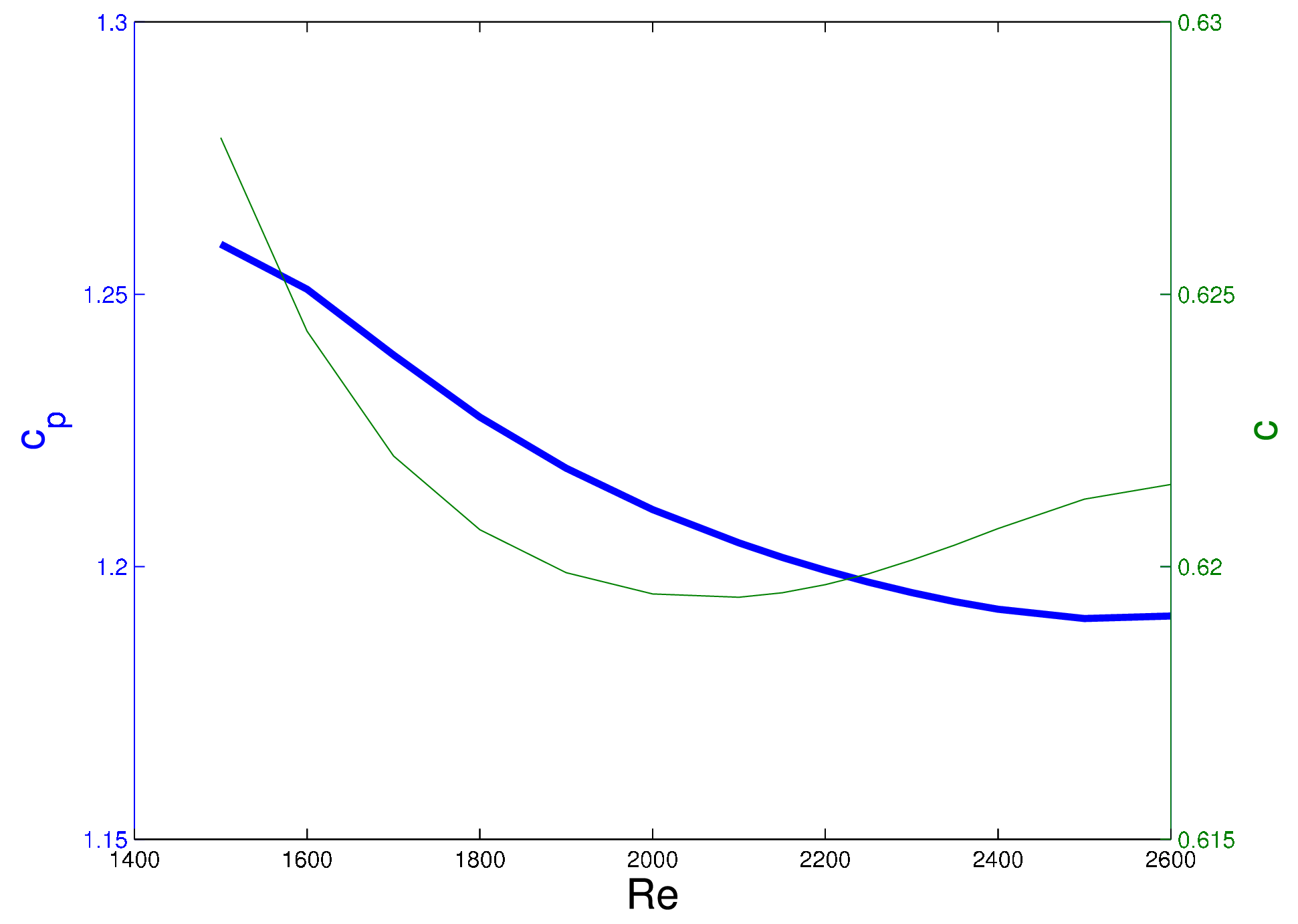}
\includegraphics[scale=0.3]{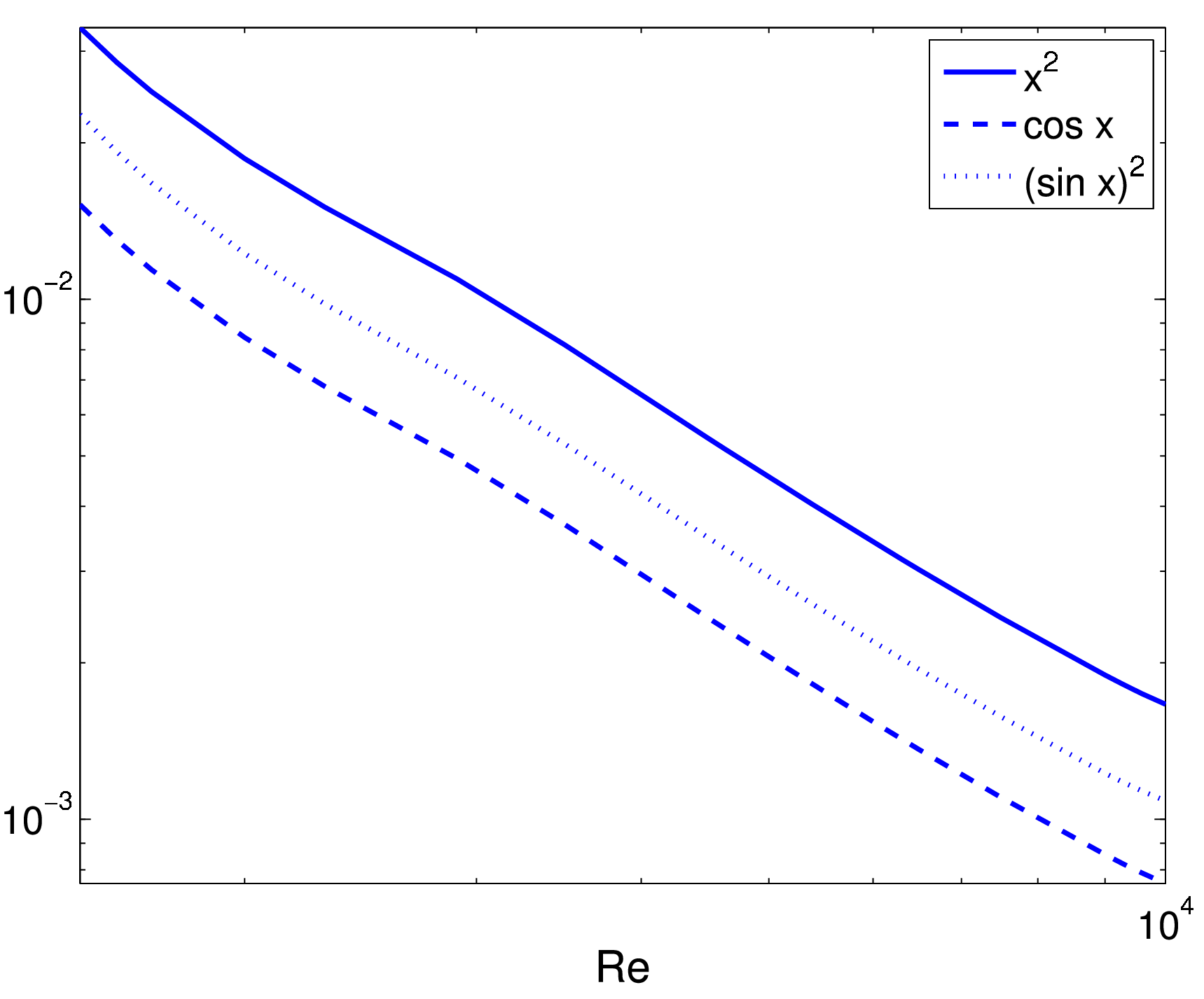}
\includegraphics[scale=0.3]{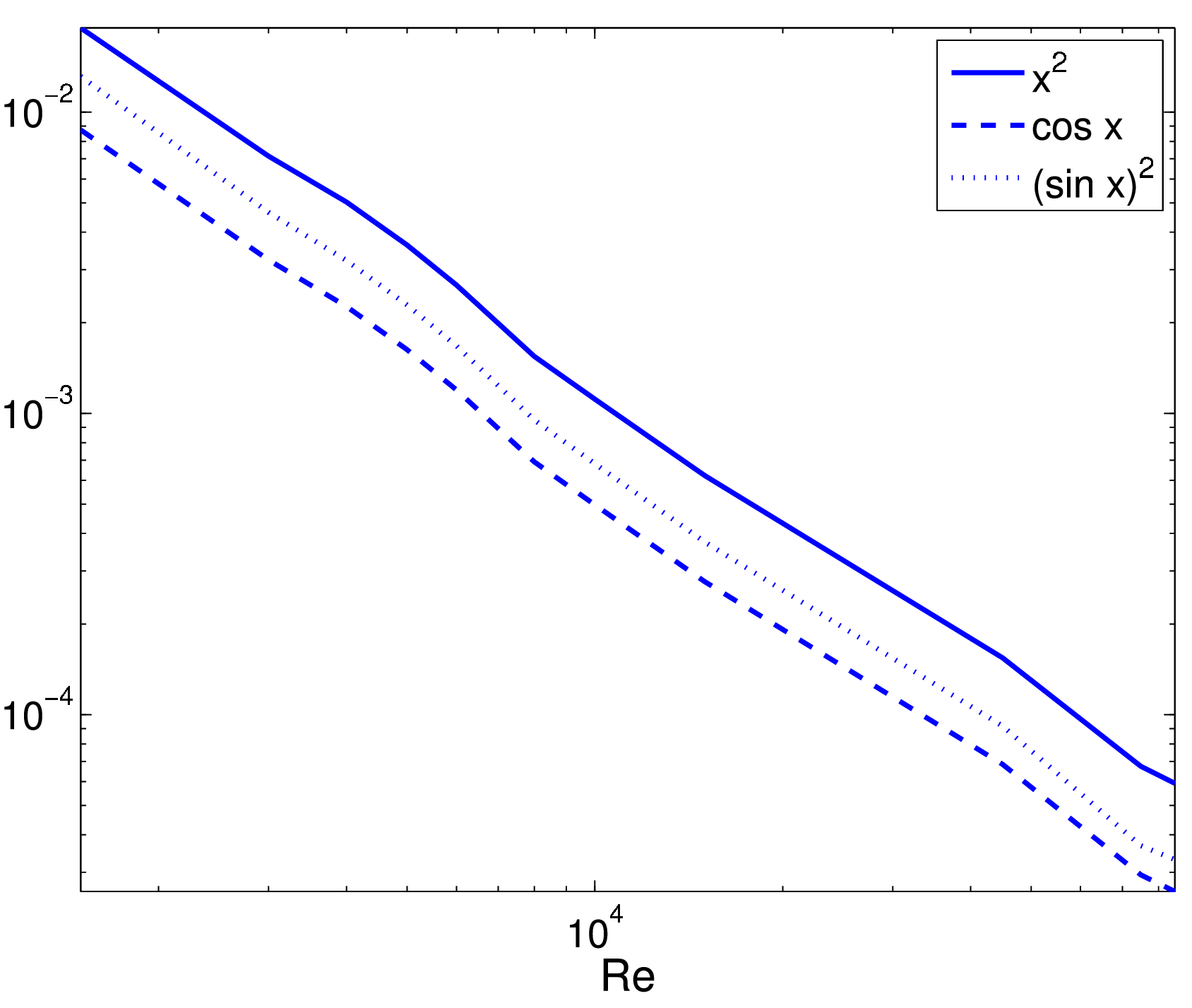}
\end{center}
\caption[xyz]{Three traveling wave solutions of pipe flow are shown.
Units are chosen to make the pipe radius and the centerline speed of
the laminar solution equal to $1$. The boundary condition fixes the
mass flux (and not the pressure gradient) to be that of the laminar
solution. Solutions (a) and (b) correspond to the $c$ family and the
upper dash-dot curve of Figure 7 of \cite{WK2004}, respectively. For
more on (c), see
\cite{Viswanath2009}. In the upper plots, the four contour levels of
$\abs{u_1}$ are equispaced in $(0, {\rm max})$, with {\rm max}
being $.0094$, $.0011$, and $.0002$, respectively. The thick line
is the critical curve. The plots below correspond to the ones
directly above. Plot (d) (the thick line is for $c_p$) can be
validated against Figure 7 of \cite{WK2004} after replacing $Re$ and
$c$ by $c_p\,Re$ and $c/c_p$, respectively. In (e) and (f), the
legends show the choice of $f(x)$ and the curves illustrate
\eqref{eqn-14} ($\alpha
\approx 1.5$).  }
\label{fig-1}
\end{figure*}

Using \eqref{eqn-9}, Waleffe et al.\cite{WGW2007} estimated that
most of the variation in $v_1$ is concentrated in a region 
around the critical curve $U=c$, with the width of that region
scaling as $Re^{-1/3}$. In the case of pipe flow, an identical
argument gives $W=c$ as the equation of the critical curve.
The top set of plots in Figure \ref{fig-1} illustrate the
critical layer in the case of pipe flow.

To derive further scalings, we consider the streamwise component
of the $n=0$ mode of NSE, which is
\begin{equation}
v_0 \partial_y U + w_0 \partial_z U = Re^{-1}(c_p + \triangle U) + M
\label{eqn-10}
\end{equation}
for plane Couette or channel flow. The pipe flow analogue is $u_0
\partial_r W + (v_0/r)\partial_\theta W = Re^{-1}(c_p + \triangle W) +
M$. In \eqref{eqn-10} and its pipe flow analogue, $-M$ equals the
$n=0$ mode of the streamwise component of $((\tilde{\bf u} - {\bf
u}_0).\nabla)(\tilde{\bf u}-{\bf u}_0)$. From here on we restrict the
derivation to plane Couette flow or to channel flow.

Since ${\bf u}_0=(U,v_0,w_0)$ has zero divergence, we can find a
function $\psi(y,z)$ such that $v_0 = \partial_z \psi$ and $w_0 =
-\partial_y\psi$. We then get
\begin{equation}
L\psi = 
Re^{-1}(c_p + \triangle U) + M,
\label{eqn-11}
\end{equation}
where  
\begin{equation}
L = (\partial_y U)\,\, \partial_z - 
(\partial_z U)\,\, \partial_y. 
\label{eqn-12}
\end{equation}
The skew-symmetry of the linear operator $L$ is the key to deducing
further scalings. The skew-symmetry of $L$ is likely to be important
in attempts to find an asymptotic theory for the critical layer.

Let $\phi(y,z)$ and $\psi(y,z)$ have $z$ periods of $2\pi\Lambda_z$ and
be sufficiently smooth. The following calculation uses
integration by parts:
\begin{align*}
&\int_0^{2\pi\Lambda}\int_{-1}^1 \phi L\psi \, dy\, dz =
\int_{-1}^1 \phi \psi U_y \bigl\lvert_{0}^{2\pi\Lambda} dy\\
&- \int_0^{2\pi\Lambda}\int_{-1}^1(\psi U_y\phi_z+\psi \phi U_{yz}) dy\, dz
- \int_0^{2\pi\Lambda} \phi \psi U_z\bigl\lvert_{-1}^{1} dz\\
&+\int_0^{2\pi\Lambda}\int_{-1}^1(\psi U_z\phi_y + \psi \phi U_{yz})dy\, dz,
\end{align*}
where the subscripts are for partial derivatives. On the right hand
side, two double integral terms cancel and the single integral terms 
are both zero because $U_y$ is periodic in $z$ and
$U_z$ is zero at the walls (from no-slip). We are left with
$-\int\int \psi L\phi\, dy\, dz$ on the right, verifying 
skew-symmetry of the operator $L$.

From direct substitution into \eqref{eqn-12}, it is evident that
$L(f(U))=0$ for {\it any} smooth $f$. Thus the functions $f(U)$ are in
the kernel of the anti-symmetric operator $L$. Since the linear system
\eqref{eqn-11} can be solved for $\psi$ (or equivalently for the
rolls), the Fredholm alternative implies that
\begin{equation*}
\int_0^{2\pi\Lambda_z} \int_{-1}^{1} f(U)\bigl(c_p + \triangle U + Re\, M\bigr)\, dy\, dz 
= 0.
\end{equation*}
For lower-branch traveling wave families with $\tilde{\bf u} - {\bf u}_0$
of magnitude $Re^{-\alpha}$ with $\alpha\approx 1$, the magnitude
of $M$ is approximately $Re^{-2}$ in the limit $Re\rightarrow\infty$.
We have
\begin{equation}
\int f(U) \Bigl(\triangle U(y,z)+c_p\Bigr) dy\,dz = O(Re^{-\alpha})
\label{eqn-13}
\end{equation}
for any smooth $f$. Here $\alpha > 1$ is possible if
there are cancellations in the integral of $M$ over the
cross-section. The analogous condition for pipe flow is
given by 
\begin{equation}
\int f(W) \Bigl(\triangle W(r,\theta)+c_p\Bigr) rdr\,d\theta 
= O(Re^{-\alpha})
\label{eqn-14}
\end{equation}
with $\alpha$ as above.

In addition to the pipe families of Figure \ref{fig-1}, we computed
a lower branch equilibrium family and a traveling wave family up to
$Re=45000$ and $Re=7000$. The (a) and (b) families of Figure \ref{fig-1}
could not be continued to $Re$ much higher than shown in the top
plots. For a given resolution, we cannot expect to find solutions
if the rolls, which diminish in magnitude with $Re$, are too small
to be detected. Even after using sufficient resolution, the 
GMRES-hookstep iterations (see \cite{Viswanath2007, Viswanath2009})
became very slow. Even though the residual error could be made
quite small, the norm of the Newton steps became quite large and
increased with iteration. Although it is uncertain if the
lower branch families exist in the $Re\rightarrow\infty$ limit,
Figure \ref{fig-1} amply demonstrates that they exist for large
enough $Re$ for the predicted scalings to hold. 

The critical curves are away from the pipe walls and have an
inward indentation where the counter-rotating vortices face each
other. Since the critical behavior is evident even for $Re=2600$,
we suspect that critical curves or surfaces may give a way to visualize
the structure of puffs in transitional pipe flow. 

In summary, we have given a number of necessary conditions for
equilibrium, traveling wave, periodic, and relative periodic solutions
in plane Couette, channel and pipe geometries. We have argued for the
importance of critical layers in high $Re$ fluid flow and shown the
connection of our analysis to critical layers. In addition, 
the conditions that
we have derived are likely to be useful in the study of transitional
structures such as puffs in pipe flow.

\bibliography{paper}

\bibliographystyle{apsrev}

\end{document}